\documentclass[11pt]{article} 
\usepackage{graphicx,amstext,amssymb}

\begin{document}

\author{M.S. Borysova$^{1}$, Yu.M. Sinyukov$^{2}$, S.V. Akkelin$^{2}$, B. Erazmus$^{3}$, Iu.A. Karpenko$^{1,2}$ }
\title{Hydrodynamic source  with continuous emission in Au+Au collisions at $\sqrt{s}=200$ GeV
}
 \maketitle

\begin{abstract}
 We analyze single particle momentum spectra and
interferometry radii in central Au+Au collisions at RHIC within a
hydro-inspired parametrization accounting for continuous hadron
emission through the whole lifetime of hydrodynamically expanding
fireball. We found that a satisfactory description of the data is
achieved for a physically reasonable set of parameters when the
emission from non space-like sectors of the enclosed freeze-out
hypersurface is fairly long: $ 9$ fm/c. This protracted surface
emission is compensated in outward interferometry radii by
positive $r_{out} - t$ correlations that are the result of an
intensive transverse expansion. The main features of the
experimental data are reproduced: in particular, the obtained
ratio of the outward to sideward interferometry radii is less than
unity and decreases with increasing transverse momenta of pion
pairs.  The extracted value of the temperature of emission from
the surface of hydrodynamic tube approximately coincides with one
found at chemical freeze-out in RHIC Au+Au collisions.
 A significant contribution of the surface emission
to the spectra and to the correlation functions at relatively
large transverse momenta should be taken into account in advanced
hydrodynamic models of ultrarelativistic nucleus-nucleus
collisions.
\end{abstract}

\begin{center}
{\small \textit{$^{1}$ Taras Shevchenko National University, Kiev 01033, Volodymirs'ka 64, Ukraine. \\[0pt]
}} {\small \textit{$^{2}$ Bogolyubov Institute for Theoretical
Physics, Kiev
03143, Metrologichna 14b, Ukraine. \\[0pt]
}} {\small \textit{$^{3}$ SUBATECH, (UMR, Universite, Ecole des
Mines, IN2P3/CNRS ), 4, rue Alfred Castler, F-44070 Nantes Cedex
03, France. \\[0pt]
}}


\end{center}

\section{Introduction}
    It is quite possible that a new state of matter is created at
unprecedented high energy density reached in relativistic
heavy-ion collisions at RHIC  \cite{RHIC}. This conclusion is
based, in particular, on the success of hydrodynamic models in
description of the measured single-particle hadron momentum
spectra and elliptic flows \cite{Heinz}. Within the hydrodynamic
approach, there is, in principle, a possibility to extract the
equation of state (EoS) of the thermalized system and then get
information about the early stage of the collision processes and
even the type of the phase transition between QGP and the hadron
matter.

However  quantitative determination of the EoS in the hydrodynamic
approach is a nontrivial problem. It depends on both the initial
conditions and  system "freeze-out" prescription. The latter
determines the final conditions of hydrodynamic expansion: one
cannot use hydrodynamic equations for infinitely large times since
the resulting very small densities destroy the picture of
continuous medium of real particles. Usually, the freeze-out
hypersurface which confines the 4-volume of hydrodynamic evolution
is included as an external input with respect to hydrodynamic
equations. Then, typically, particle spectra are described by the
Cooper-Frye prescription (CFp) \cite{Cooper} which treats
freeze-out as a sudden transition from local thermal equilibrium
to free streaming that happens on some space-like 3D hypersurface,
e.g., on a space-like part of the isotherm: $T\simeq m_{\pi}$
 \cite{Landau}.
  In reality the process of freeze-out, or particle liberation, is quite
complicated because the particles escape from the system during
the whole period of its evolution (see, e.g.,  \cite{Bravina}). A
method which is based on Boltzmann equation was recently proposed
in Ref. \cite{AkkSin1} to describe the spectra formation in the
hydrodynamic approach to A+A collisions using the escape
probabilities. The region of applicability of CFp and its possible
generalizations are discussed there, as well as in Ref.
\cite{AkkSin2}. In present work we do not consider the whole
complexity of the freeze-out process (see, e.g., Refs.
\cite{AkkSin1,AkkSin2,Bugaev1,Csern}). Instead we account for
continuous (in time) emission process by applying the generalized
Cooper-Frye prescription for an \textit{enclosed} freeze-out
hypersurface. Such an enclosed hypersurface contains space-like
sectors of the volume emission and non-space-like ones (with
space-like normal) of the surface emission.

It is well known \cite{Pratt} that protracted surface emission may
lead to the ratio of the outward to sideward interferometry radii,
$R_{out}/R_{side}$, much bigger than unity for typical scenarios
of phase transition in heavy ion collisions  \cite{Rischke}.
  These expectations are, however,  in contradiction with current experimental data
from RHIC where $R_{out}/R_{side}\simeq 1$ (see, e.g.,
\cite{ratio-exper}). This discrepancy is a component of the
so-called "HBT puzzle" \cite{puzzle}. Naively, from this one might
conclude that the duration of emission is very short.  However,
the $R_{out}/R_{side}$ ratio is also quite sensitive to the shape
of freeze-out hypersurface because, unlike $R_{side}$, $R_{out}$
is a mixture of the \textit{out-} width, time spread of emission,
and the $r_{out} - t$ correlations. Indeed, in the Gaussian
approximation for the correlation function the interferometry
radii can be expressed in terms of the space-time
variances\footnote{Note that if the distribution function has a
non-Gaussian form, which is typical for expanding sources, then
the expression of the interferometry radii through space-time
variances can be associated with the behavior of the correlation
function at small $q=p_{1}-p_{2}\rightarrow 0$ only, and,
therefore, could be used solely for qualitative estimates.} (see,
e.g., \cite{Heinz-rev}):
\begin{eqnarray}
R_{i}^{2}(p)=\langle (\triangle r_{i}-v_{i}\triangle
t)^{2}\rangle_{p}=\langle\triangle
r_{i}^2\rangle_{p}+v_{i}^2\langle\triangle t^2\rangle_{p}
-2v_{i}\langle\triangle r_{i}\triangle t\rangle_{p},
\label{variance}
\end{eqnarray}
where  $v_{i}=p_{i}/p_{0}$, $p_{i}$, $r_{i}$ ($i$ = $out$, $side$,
$long$)  are the Cartesian components of the vectors $\mathbf{v}$,
$\mathbf{p}$, and $\mathbf{r}$, respectively, and
 $p^{\mu}=(p_{0},p_{out},0,p_{long})$ is the
mean 4-momentum of  the two registered particles. Here  $\triangle
r_{i}=r_{i}-\langle r_{i} \rangle_{p}$,  $\triangle t= t-\langle t
\rangle_{p}$, and $\langle ... \rangle_{p}$ denotes  the averaged
(over the distribution function) value taken at some momentum $p$.
Note that in the Bertsch-Pratt  reference frame \cite{frame}
$p_{side}=0$ and therefore $p_{out}=p_{T}$, where $p_{T}$ is the
absolute value of the transverse component of the vector
$\mathbf{p}$.
 It is easy to see from Eq.
(\ref{variance}) that \textit{positive} $r_{out} - t$
correlations, $\langle\triangle r_{out}\triangle t\rangle_{p}>0$,
give a \textit{negative} contribution to $R_{out}$ reducing
thereby the $R_{out}/R_{side}$ ratio.\footnote{This mechanism of
reduction of the $R_{out}/R_{side}$ ratio is realized  in a
multiphase transport (AMPT) model \cite{Ko}.} Therefore one can
conclude that relatively small $R_{out}/R_{side}\simeq 1$ ratio in
a case of a prolonged surface emission can take place if there are
positive $r_{out} - t$ correlations in the corresponding sector of
the freeze-out hypersurface. Such a sector has typically a
space-like 4-normal. Then an interval between space-time points in
such a sector can be, generally, space-like as well as time-like.
While the bulk of hadrons are emitted from the sector with a
time-like normal (volume emission), the surface emission from the
sector with a space-like normal can be, nevertheless, important
for spectra and HBT radii formation\footnote{See the relevant
analysis in Ref. \cite{Heiselberg}}, especially at relatively high
$p_{T}$.
  A difference in temperature values and intensities of collective flows\footnote{It
was found in Ref. \cite{Molnar} that the transport freeze-out
process is similar to evaporation: high-$p_{T}$ particles freeze
out early from the surface, while low-$p_{T}$ ones decouple later
from the system's center.} in  different freeze-out sectors can
increase the effect, and thus a consistent description of the
single-particle spectra and interferometry radii can be reached in
hydrodynamically-motivated models with an enclosed freeze-out
hypersurface that accounts  for continuous (in time) character of
particle emission.

 It is well known that when
the freeze-out hypersurface  $\sigma$ has non-space-like sectors,
the CFp is inconsistent. Then the CFp should be modified to
exclude formally negative contributions to the particle number at
the corresponding momenta.  The simplest prescription is to
present the distribution function as a product of a local thermal
distribution and the step function  like $\theta (p_{\mu }n^{\mu
}(x))$ \cite{Sin4,Bugaev}, $ n^{\mu }n_{\mu }=\pm 1$, where $
n^{\mu }$ is a time-like or space-like outward normal to a
freeze-out hypersurface $\sigma $.\footnote{If a fluid element
that crosses a sector of the freeze-out hypersurface with a
space-like normal decays, preserving its total particle number,
then this prescription is somewhat more complicated \cite{Sin4}.}
Thereby freeze-out is restricted to those particles for which
$p_{\mu }n^{\mu }(x)$ is positive.

 Sometimes, to avoid the above mentioned problems with
CFp, the continuity of particles emission in heavy ion collisions
is taken into account  by means of the emission (source) function
$S(x,p)$  (see, e.g., \cite{Heinz-rev}), which is used instead of
the distribution (Wigner) function $f(x,p)$ in some
 hydro-inspired parameterizations and is usually chosen  to be proportional to the local
 equilibrium distribution function, $f_{l.eq}(x,p)$, with a smearing (proper) time factor
 $\exp(-(\tau-\tau_{0})^2/\Delta \tau^{2})$.
 First, such a prescription loses an extremely important
 information about $\tau-r_{i}$ correlations:
 it is naturally that at early times the emission function is concentrated mostly at the  periphery
 of the system on the boundary with vacuum, since particles cannot escape from the hot and dense central region.
 The approximation of surface emission is, thus, much more realistic for the early times. Second,
while the distribution function is well defined for a given state
of the system, like the local Bose-Einstein or Fermi-Dirac
distributions are associated with a locally equilibrated state,
$S(x,p)$ is much more complicated and model dependent
\cite{AkkSin2,Zalewski}. It is demonstrated in Ref.
\cite{AkkSin1}, starting from a particular exact solution of the
Boltzmann equation, that the emission function is not proportional
to the local equilibrium distribution (Wigner function), although
the system is in a local equilibrium state. Therefore the widely
used (see, e.g., so called "blast-wave" model \cite{blast}) "local
equilibrium" ansatz, $S\sim f_{l.eq.}$,
 is supported neither by analytic kinetic results nor numerical transport
calculations except for the formal case when all the particles are
emitted simultaneously: the emission function is then the
distribution function multiplied by a delta function concentrated
on the freeze-out hypersurface. Perhaps, too large values of
maximal collective velocities, close to the speed of light, which
are necessary to fit the spectra in the blast-wave
parametrization, and a failure in fitting HBT radii (especially
$R_{side}$, see Refs. \cite{blast}) are caused by an improper
description of the particle emission process.

The aim of this work is a coherent description of pion, kaon and
proton single-particle momentum spectra as well as the HBT pion
radii in central RHIC Au+Au  collisions at $\sqrt{s_{NN}}=200$ GeV
by means of a simple hydrodynamically-motivated parametrization
with an enclosed freeze-out hypersurface. As a matter of fact, we
generalize the blast-wave parametrization for a realistic case of
continuous emission over the whole lifetime of the fireball.
  We demonstrate that a consistent
description of the observables is possible if the freeze-out
hypersurface has a non-space-like sector emitting a noticeable
part of the particles.\footnote{The importance of surface emission
has been stressed also in Refs. \cite{opaque} for opaque sources
which emit particles from a surface layer of finite thickness,
then $\langle\triangle r_{out}^2\rangle_{p}< \langle\triangle
r_{side}^2\rangle_{p}$ reducing thereby $R_{out}/R_{side}$ ratio.}
The latter condition is satisfied when particles are emitted from
the surface of an expanding hydrodynamic tube soon after the
hadronization, which at early times happens at the periphery of
the volume occupied by quark-gluon matter.

\section{Hydro-inspired parametrization of continuous hadronic freeze-out in central RHIC Au+Au collisions}
 We  assume that
particles are emitted from some 3D hypersurface when the fireball
is in a locally equilibrated (leq) state,
\begin{equation}
f_i=f_{leq,i}(x,p)=(2\pi)^{-3} \left (\exp \left
(\frac{u_{\nu}(x)p^{\nu}-\mu_{i} (x) }{T(x)} \right ) \pm 1 \right
)^{-1}, \label{f-l.eq.}
\end{equation}
characterized by the  temperature $T(x)$ (which is common for all
the particle species), collective velocity field $u_{\mu}(x)$ and
the chemical potentials $\mu_{i}(x)$ depending on the particle
species $i$. We refer to the part of 3D hypersurface with a
time-like normal vector as $v$- (volume) freeze-out, or final
breakup, and the rest of the hypersurface with a space-like normal
as $s$- (surface) freeze-out, corresponding to continuous emission
prior to the final breakup. We assume that particle densities are
uniform at each $v-$ and $s-$ hypersurfaces separately, but, in
general, are different from each other. Then the temperature and
chemical potential in
 Eq. (\ref{f-l.eq.}) are replaced with $T_{v}$, $T_{s}$ and
$\mu_{v,i}$, $\mu_{s,i}$ for the $v-$ and $s-$ freeze-out sectors,
respectively. For the sake of simplicity and reduction of the
number of parameters we do not interpolate smoothly the values of
temperature and chemical potential between the \textit{volume} and
\textit{surface} sectors of the hypersurface. A longitudinally
boost-invariant expansion is a rather good approximation in
mid-rapidity region at RHIC \cite{Heinz-hyd}, therefore we assume
boost invariance for longitudinal hydrodynamic velocities in this
region, $v_{L}=z/t$, which allows us to parameterize $t$ and $z$
at 3D hypersurface $ \sigma $ as $t_{v,s}=\tau_{v,s} (r)\cosh
(\eta )$, $z_{v,s}=\tau_{v,s} (r)\sinh (\eta )$ in both sectors of
the freeze-out hypersurface, where $\tau=\sqrt{t^{2}-z^{2}}$ is
the Bjorken proper time \cite{Bjorken} and $\eta$ is the
longitudinal fluid rapidity, $\eta =\tanh ^{-1}v_{L}$; in
addition, $r$ is the absolute value of transverse coordinate ${\bf
{r_{T}}}\equiv (x,y)=(r\cos \phi,r\sin \phi)$. Taking into account
the transverse velocity component $v(r,\tau )$ in the
longitudinally co-moving frame, we obtain for the collective
4-velocity
\begin{equation}
u^{\mu }_{v,s}(r,\eta )=( \cosh \eta \cosh \eta_{T}^{v,s}, \sinh
\eta_{T}^{v,s} \cos \phi , \sinh \eta_{T}^{v,s} \sin \phi ,\sinh
\eta \cosh \eta_{T}^{v,s}) , \label{transv}
\end{equation}
where $\eta_{T}^{v,s}(r,\tau_{v,s})$ is the transverse fluid
rapidity, $\eta_{T} =\tanh ^{-1}v(r,\tau )$. The particle
4-momentum can be expressed through the momentum rapidity $y$,
transverse momentum $\mathbf{p_{T}}$ and transverse mass
$m_{T}=\sqrt{m^{2} + p_{T}^{2}}$,
\begin{equation}
p^{\mu}=(m_{T}\cosh y, \mathbf{p_{T}}, m_{T}\sinh y).
\label{part-def1}
\end{equation}
Then
\begin{equation}
\frac{p_{\mu}u^{\mu }}{T}=\frac{ m_{T}}{T}\cosh{(y-\eta)} \cosh
{\eta_{T}} -
\frac{\mathbf{p_{T}}}{T}\frac{\mathbf{r_{T}}}{r}\sinh{\eta_{T}}.
\label{p-u}
\end{equation}

The \textit{volume } and \textit{surface } elements of the
freeze-out hypersurface $\sigma (x)$ take the form
\begin{equation}
d\sigma _{\mu }^{v,s}=n_{\mu}^{v,s}d\sigma=\pm \tau_{v,s} (r)d\eta
dr_{x}dr_{y}( \cosh \eta ,-d\tau_{v,s} /d{r}_{x},-d\tau_{v,s}
/d{r}_{y},-\sinh \eta ), \label{sigm}
\end{equation}
where
\begin{equation}
n_{\mu}^{v,s}=\pm\frac{( \cosh \eta ,-d\tau_{v,s}
/d{r}_{x},-d\tau_{v,s} /d{r}_{y},-\sinh \eta
)}{(\pm(1-(d\tau_{v,s} /d{r}_{x})^{2}-(d\tau_{v,s}
/d{r}_{y})^{2}))^{1/2}} . \label{vector}
\end{equation}
and "+" and "-" correspond to the $v-$ and $s-$  sectors of the
hypersurface respectively.\footnote{Note that the above
representation of elements of the freeze-out hypersurface $\sigma
(x)$ in the s-sector is, of course, not unique and is used because
of its convenience for the chosen type of $\sigma (x)$ (see
below). It should be changed if $ dr_{s}(\tau)/d\tau =0 $ on some
part of this hypersurface.} The surface "moves" with a velocity
less than the speed of light, $ |dr_{s}(\tau)/d\tau| <1 $, hence
$|d\tau_{s}(r)/dr|>1$. If the 2D surface of the 3D $s-$ freeze-out
sector "moves" outwards, $d\tau_{s} /d{r}_{x},d\tau_{s}
/d{r}_{y}>0$, then the product $p^{\mu}n_{\mu}$ will be negative
if the $p_{T}$ of particles are sufficiently small. In that case
such particles cannot really escape from the system. For particles
with fairly high transverse momenta one has $p^{\mu}n_{\mu}>0$,
and they will be emitted from the surface of the system into
vacuum and stream freely towards detectors. To take this into
account we use a modified Cooper-Frye prescription with the
substitution \cite{Sin4,Bugaev}
\begin{equation}
f_{l.eq.}(x,p)\rightarrow \theta (p_{\mu }n^{\mu
}(x))f_{l.eq.}(x,p) \label{theta}
\end{equation}
to get the single particle spectra $p_{0}d^{3}N/d^{3}p$  for
diverse particle species  and the correlation function of pions
$C(p,q)$:
\begin{eqnarray}
p_{0}\frac{d^{3}N}{d^{3}p}=\int p^{\mu}d\sigma_{\mu} \theta
(p_{\mu }n^{\mu}(x))f_{l.eq.}(x,p)  \label{sp-def} \\
C(p,q)= 1+\frac{|\int p^{\mu}d\sigma_{\mu} \theta (p_{\mu
}n^{\mu}(x))f_{l.eq.}(x,p)\exp(iqx)|^{2}}{\left(\int
p_{1}^{\mu}d\sigma_{\mu} \theta (p_{1}^{\mu
}n_{\mu}(x))f_{l.eq.}(x,p_{1})\right)\left(\int
p_{2}^{\mu}d\sigma_{\mu} \theta (p_{2}^{\mu
}n_{\mu}(x))f_{l.eq.}(x,p_{2})\right)}.
  \label{corr-def}
\end{eqnarray}
Here  $p=(p_{1}+p_{2})/2$, $q=p_{1}-p_{2}$. To calculate the
correlation function in the region of correlation peak we utilize
 the smoothness mass shell approximation:
\begin{equation}
C(p,q)\approx 1+\frac{|\int p^{*\mu}d\sigma_{\mu} \theta
(p^{*}_{\mu
}n^{\mu}(x))f_{l.eq.}(x,p^{*})\exp(iqx)|^{2}}{\left(\int
p^{*\mu}d\sigma_{\mu} \theta (p^{*\mu
}n_{\mu}(x))f_{l.eq.}(x,p^{*})\right)^{2}} \label{corr-approx}
\end{equation}
where
$p^{*\mu}=(\sqrt{m^2+(\frac{\mathbf{p_{1}}+\mathbf{p_{2}}}{2})^2},\frac{\mathbf{p_{1}}+\mathbf{p_{2}}}{2})$.

We assume that the source of particles in central Au + Au RHIC
collisions has cylindrical symmetry and is strongly stretched out
in the beam direction ($long-$ direction): the longitudinal size
is much larger than the corresponding length of homogeneity
\cite{Sinyukov,A-B-S}.  The latter allows us to neglect the
finiteness of the system in the longitudinal direction when we
calculate particle spectra in the mid-rapidity region around $y
\approx 0$. As for the transverse direction, we assume that the
source has a finite geometrical size encoded in the limits of
integration over $r$: $0<r<R_{f}$ for $v-$emission and
$R_{i}<r<R_{f}$ for $s-$emission, where $R_{i}$ and $R_{f}$ are
the initial and finite effective radii of the system.

 To fix a model one needs to define $\tau (r)$ and the transverse
rapidity $\eta_{T}(r)$. We choose the simplest ansatz,  aimed at
catching the main features of particle emission.
 We suppose that the \textit{volume} (space-like)  emission happens as assumed
in the blast-wave model at the constant $\tau_{v}(r)=\tau_{f}$,
and the \textit{surface} emission takes place during the whole
time of hydrodynamic evolution of system:
$\tau_{i}<\tau_{s}<\tau_{f}$, where $\tau_{i}$ is the initial time
of thermalization. Because the fireballs formed in RHIC collisions
expand in transverse direction with rather high transverse
velocity, we can assume that $d\tau_{s}(r)/dr>0$ for the
\textit{surface }freeze-out.\footnote{This is in full
correspondence with the form of the hypersurface of constant
densities (particle number and energy) in analytic ellipsoidal
solutions of relativistic hydrodynamics \cite{Karpenko}. It seems
that initial or fast developed transverse velocities are required
in order to provide a positive $t-r$ correlation in {\textit s}-
sector of freeze-out hypersurface.} For the sake of simplicity we
parameterize $\tau_{s}(r)$ by a linear function,
\begin{equation}
\tau_{s}(r)=a \cdot r + b. \label{tau-s}
\end{equation}
Taking into account that $\tau_{s}(R_{i})=\tau_{i}$ and
$\tau_{s}(R_{f})=\tau_{f}$, where $R_{i}$ and $R_{f}$ are the
initial and final transverse radii respectively, $R_{i}<R_{f}$, we
obtain that
\begin{eqnarray}
a=\frac{\tau_{f}-\tau_{i}}{R_{f}-R_{i}},  \label{a} \\
b=\tau_{i} - \frac{\tau_{f}-\tau_{i}}{R_{f}-R_{i}}R_{i}.
  \label{b}
\end{eqnarray}
Please note that the chosen form of $\tau_{s}(r)$ is just an
approximation that reflects the main properties of the
\textit{surface} freeze-out. For example, the chosen constant
velocity of outward motion of the transverse surface, $1/a$, is
associated with the mean (averaged over time) velocity of the
surface, which generally can move  with time-dependent velocity.

The form of freeze-out $\tau (r)$ is presented in Fig. 1. It is
convenient to parameterize the \textit{volume} as well as
\textit{surface} sectors of the 3D hypersurface by
$\mathbf{r}_{T}$ and longitudinal rapidity $\eta$, therefore  the
integration   in Eqs. (\ref{sp-def}), (\ref{corr-approx}) is
carried out separately for those sectors: $\int d\sigma _{\mu
}[...]\rightarrow \int d\sigma _{\mu }^{v}[..]_{v}+\int d\sigma
_{\mu }^{s}[..]_{s}$.

To specify the transverse rapidity $\eta_{T}(r)$, we first connect
smoothly the values of transverse rapidity in the \textit{volume}
and \textit{surface }sectors of the hypersurface on their
"border". Second, since a linear $r$ dependence of the radial
velocity profile was found in hydrodynamic simulations \cite{Kolb}
for small transverse collective velocities $v \approx \eta_{T}$
near the center of the system at fixed $\tau$, we assume that the
transverse flow rapidity profile depends linearly on $r$ for
\textit{volume} freeze-out at $\tau = \tau_{f}$,
\begin{equation}
\eta_{T}^{v}(r,\tau_{f})=\eta_{T}^{max}\frac{r}{R_f},
\label{eta-v}
\end{equation}
 where $\eta_{T}^{max}$   is the maximal transverse rapidity of the fluid.
  As for the transverse rapidity profile in the \textit{surface } sector of the freeze-out hypersurface,
 we analyzed various  parameterizations.  Fits to the data
 prefer a non-zero initial transverse velocity at $\tau _{i} = 1$ fm/c, which is physically
reasonable. Indeed, it seems to be natural that the transverse
flow velocity at freeze-out surface at any time (or, at least, on
average) should not be smaller than the average "velocity" of the
surface, $1/a$. Such an  "initial" velocity, $ v_{T}(\tau _{i} =
1$ fm/c), can be developed even at the non-thermal partonic stage
before $\tau _{i}$. Also a non-zero initial velocity does not
contradict an approximate linearity with radius of rapidity
profile at early times, in particular at
 $\tau_{i}$. One of the simplest parameterizations which satisfies these properties  and allows us
 to connect smoothly the values of
 transverse rapidity in the \textit{volume} and \textit{surface} sectors is the following
 transverse rapidity distribution:
\begin{equation}
\eta_{T}^{s}(r,\tau_{s}(r))=\eta_{T}^{max}\frac{\sqrt{(\tau_{s}(r)-\tau_{i})^2+r^{2}}}
{\sqrt{(\tau_{f}-\tau_{i})^2+R_{f}^{2}}}.  \label{eta-s}
\end{equation}
Note that $\tau_{s}(R_{f})=\tau_{f}$ and
$\eta_{T}^{s}(R_{f},\tau_{s}(R_{f}))=\eta_{T}^{v}(R_{f},\tau_{f})=\eta_{T}^{max}$.

   In the next Section we apply our model to description of pion, kaon and proton transverse momentum spectra
and pion interferometry radii in mid-rapidity in RHIC Au+Au central
collisions at $\sqrt{s_{NN}}=200$ GeV.

\section{Results and discussion}
We assume a common enclosed freeze-out hypersurface for all
particles species. Then the model contains 7 parameters:
$\tau_{i}$, $\tau_{f}$, $R_{i}$, $R_{f}$, $T_{v}$, $T_{s}$,
$\eta_{T}^{max}$. As commonly accepted, we assume that at
$\tau_{i}=1$ fm/c the system is already in local thermal
equilibrium, and so we will not consider this value as a free
parameter.

The values of chemical potentials $\mu_{s,j}$, $\mu_{v,j}$ which
are different for different particle species $j$, are not free
parameters and are responsible for the absolute normalization of
the spectrum.  In details,  the situation is somewhat more
complicated since resonance decays contribute to both the particle
spectra and interferometry radii \cite{Padula}. About $1/2 - 2/3$
of pions, kaons and protons come from those decays after chemical
freeze-out happens in high-energy heavy ion collisions
\cite{Braun-Munzinger,PBM-1,A-B-S}. The corresponding contribution
to pion spectra and interferometry radii was estimated in Ref.
\cite{AkkSin} within the hydrodynamic model developed in
\cite{A-B-S}. According to the results of Ref. \cite{AkkSin},
resonance decays do not change significantly the slopes of the
transverse pion spectra in the region of interest for kinetic
freeze-out temperatures $T_{v}$ and flows at RHIC collisions. Note
that the resonance emission from the surface is negligible since
heavy resonances have too small heat velocities to escape from the
expanding system. As for the pion interferometry radii, the decays
of short-lived resonances result in an augmentation of them as
compared with the picture of pure thermal emission, especially in
the region of small $p_{T}$. This resonance contribution to the
pion interferometry radii decreases with the increase of
transverse momenta of the pairs and, as it was found in Ref.
\cite{AkkSin}, is smaller than $6 - 14$ $\%$ for $p_{T}>0.3$ GeV
(see also \cite{Heinz-res}).

 Based on these studies, we neglect
the changes of slopes of particle transverse spectra due to
resonance decays and fit the pion interferometry radii starting
from $m_{T}\simeq 0.3$ GeV where the resonance contributions are
fairly small. Finally, to fix $\mu_{v,j}$, $\mu_{s,j}$ for thermal
(primary) particles at some $T_{v}$ and $T_{s}$, one needs to know
what fraction of measured particles is produced by resonances
after thermal (kinetic) freeze-out.
 Such a problem is not fully solved until now and needs
further analysis. Since the main aim of this paper is to describe
the main features of the physical picture of particle emission in
central RHIC Au+Au collisions rather than to fine tune the
parameters and perfectly describe the data, we fit the fraction of
thermal (direct) particles within reasonable limits discussed
above; our final choice is presented in the last column of Table
1.

A preliminary analysis shows that a good fit to spectra and
correlations formed on an enclosed hypersurface cannot be obtained
if the \textit{surface} temperature $T_{s}$ is taken within the
standard range for kinetic freeze-out: $100$ MeV $\leq T_{s}\leq
m_{\pi}$; it should be taken higher than the pion mass
$m_{\pi}$.\footnote{The same conclusion that the best fit of the
pion HBT radii is obtained when the  temperature of
\textit{surface} emission from a hypersurface with a spacelike
normal is significantly higher  than the temperature of
\textit{volume} emission is made also in Ref. \cite{Renk}, where
SPS data for 158 AGeV central Pb+Pb collisions are analyzed.} It
forces us to conclude that the temperature of surface emission is
close to the chemical freeze-out temperature, or the hadronization
one if chemical freeze-out happens at about the same time, as it
is argued in Ref. \cite{PBM}. Since the periphery of an expanding
system is more dilute, the matter there is in the hadronic phase
and hadrons can easily escape from this layer into the surrounding
vacuum despite the relatively high temperature; this was argued in
detail in Ref. \cite{Sin4}.  This natural picture of surface
emission at earlier times $\tau_{s}<\tau_{f}$ gives us a
possibility to assume $\mu_{s}=0$ for pions and kaons, accounting
for chemical equilibrium of pions at chemical freeze-out (see,
e.g., \cite{PBM-1}) and almost zero net strangeness in central
RHIC Au+Au collisions at the top energy. The results of our fit,
which are presented in Tables 1 and 2, demonstrate that the best
fit fixes the value of $T_{s}$ at $150$ MeV, which is indeed very
close to the chemical freeze-out temperature $157\pm 3$ MeV
obtained in Ref. \cite{chemic-200}. This justifies our assumption
\textit{a posteriori}. The best fit to proton transverse spectra
is achieved at $\mu_{s,p}=40$ MeV, which is slightly more than the
value $\mu_{B}=28.2\pm 3.6$ MeV found in Ref. \cite{chemic-200}
for the baryochemical potential.\footnote{Note  that it is not
only the absolute values but also the spectrum slopes that are
affected by the values of $\mu_{s,j}$ because chemical potentials
determine what part of particles are emitted from the
\textit{surface}, and these particles have, of course, different
spectrum slopes than the ones emitted from the \textit{volume}
part of the freeze-out hypersurface.} Concerning the values
$\mu_{v,j}$, one readily sees that they are not free parameters of
the model, since they should be fixed by the measured respective
rapidity densities, $dN_{j}/dy$.

The best fit to data is obtained for the parameters shown in
Tables 1 and 2. We also present there parameters that were kept
constant through the fit procedure.  As one can see from Figs. $2
- 4$ and Tables 1, 2, that good description of pion, proton and
kaon spectra and pion interferometry radii is achieved in the
mid-rapidity region of central $\sqrt{s_{NN}}=200$ GeV Au+Au
collisions. The STAR collaboration data are presented for
illustration. We did not fit these data because they are limited
to relatively small $p_T$ values as compared to the PHENIX data,
whereas the surface emission gives a significant contribution in
the region of high transverse momenta. It is noteworthy, that the
best fit is achieved when the system already at $\tau_{i}=1$ fm/c
expands in the transverse direction with a collective velocity
field that reaches at the system border (at $r=R_{i}$) the value
$v(R_{i},\tau_{i})\approx 0.34$ c. It may indicate that a
noticeable contribution into buildup of transverse flow at RHIC
energies is given at times earlier then 1 fm/c, possibly
pre-thermal stages of evolution. It is interesting to note that
hydrodynamic simulations \cite{Rapp} of heavy ion collisions at
RHIC energies also indicate the necessity of an initial
(pre-hydrodynamic) transverse flow to better account for slopes of
the observed spectra.

\begin{center}
TABLE 1.

 {\small Model  parameters from fits of  Au+Au
$\sqrt{s_{NN}}=200$ GeV data on $\pi^{-}$, $K^{-}$ and $p$
single-particle spectra and $\pi^{-}$ HBT radii. The ratio of
detected particle numbers to thermal (primary) ones (last column)
as well as $\mu_{s}$ for pions and kaons were kept constant during
fit procedures.}
\bigskip

\begin{tabular}{|c|c|c|c|}
  \hline
   & $\mu_{v}$ MeV & $ \mu_{s}$ MeV & $ \frac{dN^{reg}/dy}{dN^{th}/dy}$
\\
  \hline
$\pi^{-}$       & $53$ & $0$ & $2.2$
\\
  \hline
  $K^{-}$       & $45$ & $0$ & $2.2$
  \\
  \hline
$p$  & $280$ &  40 & $3.5$
\\
  \hline
\end{tabular}
\end{center}

\begin{center}
TABLE 2.

{\small  Model  parameters from fits of Au+Au $\sqrt{s_{NN}}=200$
GeV data on $\pi^{-}$, $K^{-}$ and $p$ single-particle spectra and
$\pi^{-}$ HBT radii.  The value of $\tau_{i}$ was kept constant
through fit procedures.}
\bigskip

\begin{tabular}{|c|c|c|c|c|c|c|c|}
  \hline
 & $\tau_{i}$ fm/c& $\tau_{f}$ fm/c & $R_{i}$ fm & $ R_{f}$ fm & $ T_{v}$ MeV &
  $T_{s}$ MeV & $\eta_{T}^{max}$
\\
  \hline
$\pi^{-}$,  $K^{-}$, $p$ & 1  & $10$  & $7$ & $11.3$ & $110$ &
$150$ & $ 0.73$
\\
  \hline

\end{tabular}
\end{center}

The solid lines in Figs. $2 - 4$ demonstrate the fit according to
our model, and the dashed lines indicate the contributions to
particle spectra and interferometry radii from \textit{volume }
freeze-out solely. The latter corresponds, in fact, to blast-wave
model calculations with our \textit{volume } freeze-out parameters
and $\Delta \tau =0$. The experimental data are taken from Refs.
\cite{PHENIX-sp,STAR-rad,PHENIX-rad}. We used Boltzmann
approximation for the distribution functions  in the
single-particle spectra and interferometry radii calculations for
all particle species. One can see from Fig. 2 that \textit{surface
}emission leads to a noticeable enhancement of the single-particle
momentum spectra in high $p_{T}$ region, resulting thereby in an
increase of the inverse spectra slope, or effective temperature.
This happens because the particles with relatively low $p_{T}$ are
not emitted from the \textit{surface} part of the freeze-out
hypersurface that moves outwards  and so their contribution to the
spectra is suppressed, while the particles with sufficiently high
$p_{T}$ are liberated freely. This effect is taken into account,
as discussed in the previous Section, by means of multiplying the
distribution functions by the step function, see Eq.
(\ref{theta}), thus restricting  the freeze-out to those particles
for which the product $p^{\mu}n_{\mu}$ is positive. Note that in
the blast-wave parametrization, which takes into account only
\textit{volume } emission, such a high effective temperature can
only be obtained at the cost of a very high transverse flow
($\eta_{T}^{max}\approx 0.9$ and so $v^{max}\approx 0.7$ c, while
in our parametrization $\eta_{T}^{max}\approx 0.72$,
$v^{max}\approx 0.6$ c)
  at freeze-out \cite{blast}, which seems
less realistic and more difficult to reproduce in hydrodynamic
models of the system evolution.

The Figures $3$, $4$ show our results for the interferometry radii
of negative pions.  Note that the interferometry radii are
measured by the PHENIX Collaboration for $0-30$ $\%$ centrality
events at $\sqrt{s_{NN}}=200$ GeV \cite{PHENIX-rad}. Therefore, to
adjust the results of the PHENIX Collaboration to the most central
$0-5$ $\%$ centrality bin, we increase the corresponding
interferometry radii by a factor calculated  in accordance with
the $N_{part}$ dependence of the Bertsch-Pratt radius parameters
found in Ref. \cite{PHENIX-rad}.
 As one can see from Fig. 3, the interplay between (i) a long
emission time, $\tau_{f} - \tau_{i}$, which gives an increase of
the $R_{out}$ radius, and (ii) positive $r_{out}-t$ correlations
at the \textit{surface } part of the hypersurface where
$d\tau_{s}/dr>0$, which reduces the radius, provides a
\textit{moderate} increase of $R_{out}$ as compared with the
\textit{volume} emission at constant $\tau$.\footnote{The
conclusion that a long-lived fireball can be compatible with HBT
measurements is made also in Ref. \cite{Renk}} The $R_{side}$
radius is almost not affected by the \textit{surface}
emission.\footnote{An independence of $R_{side}$ on transport
opacity was also found in a covariant transport model
\cite{Molnar}.} As to $R_{long}$, it is less than in the
\textit{volume } emission case. The reason is that $R_{long}$ is
strongly affected by the value of the emission time $\tau$
\cite{Sinyukov} and, thereby, is reduced by the emission from
early times $\tau < \tau _{f}$. This can explain an apparently low
freeze-out value of $\tau$ in relativistic heavy ion collisions if
it is extracted assuming instantaneous \textit{volume} emission at
constant $\tau$.

 Figure 4 represents our result for the
$R_{out}/R_{side}$ ratio. One can see that the \textit{volume}
blast-wave-like emission (at a constant $\tau$) leads to a rather
low $R_{out}/R_{side}$ ratio. The "extra" reduction of $R_{out}$
is typical for a sharp freeze-out hypersurface with step density
profile in the out-direction, $n(r)\sim \theta(R-r_{out})$ and the
maximal velocity on the edge. The same boundary effect of the
additional broadening of the correlation function at large momenta
in the case of  sharp step-like freeze-out in the long-direction
have been found and described in detail in Ref.
\cite{Averchenkov}. As one can see from Fig. 4, this effect
becomes softer for an enclosed freeze-out hypersurface despite the
the fact that maximum velocity is also achieved at maximum
transverse radius which the system occupies during its evolution.
However, negative $t-r_{out}$ correlations can totally destroy the
effect of relative reduction of $R_{out}$. That is why the
positive $t-r_{out}$ correlations in our model are very important
to achieve a satisfactory description of the experimental data.

\section{Summary}
We have studied the effect of continuous particle emission on
particle spectra and interferometry radii, based on hydro-inspired
parametrization of the enclosed freeze-out hypersurface and
distribution functions. Our model generalizes the "blast-wave"
parametrization, which has difficulties in the fitting of the HBT
radii and leads to too large transverse velocities. We demonstrate
that $\pi^{-}$, $K^{-}$ and $p$ single particle spectra and
$\pi^{-}$ interferometry radii at RHIC can be described by a
physically reasonable set of parameters when a rather protracted
emission ($\sim 9$ fm/c) from the \textit{surface} sector of the
hypersurface that moves "outward" is taken into account.
 The phase-space distribution of particles
emitted from the surface prior to the fireball breakup is similar to
the local chemical equilibrium distribution with parameters rather
close to the ones found at chemical freeze-out in RHIC Au+Au
collisions. This
 could indicate a transformation of quark-gluon
matter to hadron gas at the periphery of the system with
subsequent hadronic emission just after such a transition from the
boundary of the fireball during the whole time of the system's
hydro evolution.

 Thereby, the simple physical arguments requiring the freeze-out hypersurface
to be enclosed for finite expanding systems result in a successful
description of hadronic spectra and interferometry radii. This
means that particle emission from the early stages of fireball
evolution has to be taken into account in advanced hydrodynamic
models of A+A collisions.\footnote{The back reaction of
evaporation of particles off the surface of the system created on
the fluid dynamics was studied within boost-invariant
hydrodynamics in Ref. \cite{Stocker}.}  A quantitative description
of a continuous escaping of particles can be made, in particular,
within the hydro-kinetic approach to spectra formation proposed in
Ref. \cite{AkkSin1}.

\section*{Acknowledgments}
 We are grateful to Stefan Mashkevich and Sergey Panitkin for careful
reading of the manuscript and useful suggestions. The research
described in this publication was made possible in part  by Award
No. UKP1-2613-KV-04 of the U.S. Civilian Research $\&$ Development
Foundation for the Independent States of the Former Soviet Union
(CRDF). The research was carried out within the scope of the ERG
(GDRE): Heavy ions at ultrarelativistic energies - a European
Research Group comprising IN2P3/CNRS, Ecole des Mines de Nantes,
Universite de Nantes, Warsaw University of Technology, JINR Dubna,
ITEP Moscow, and Bogolyubov Institute for Theoretical Physics, NAS
of Ukraine. The work was also supported by NATO Collaborative
Linkage Grant No. PST.CLG.980086 and Fundamental Research State
Fund of Ukraine, Agreement No. F7/209-2004.

\newpage

\begin{figure}[h]
\centering
\includegraphics[scale=1.0]{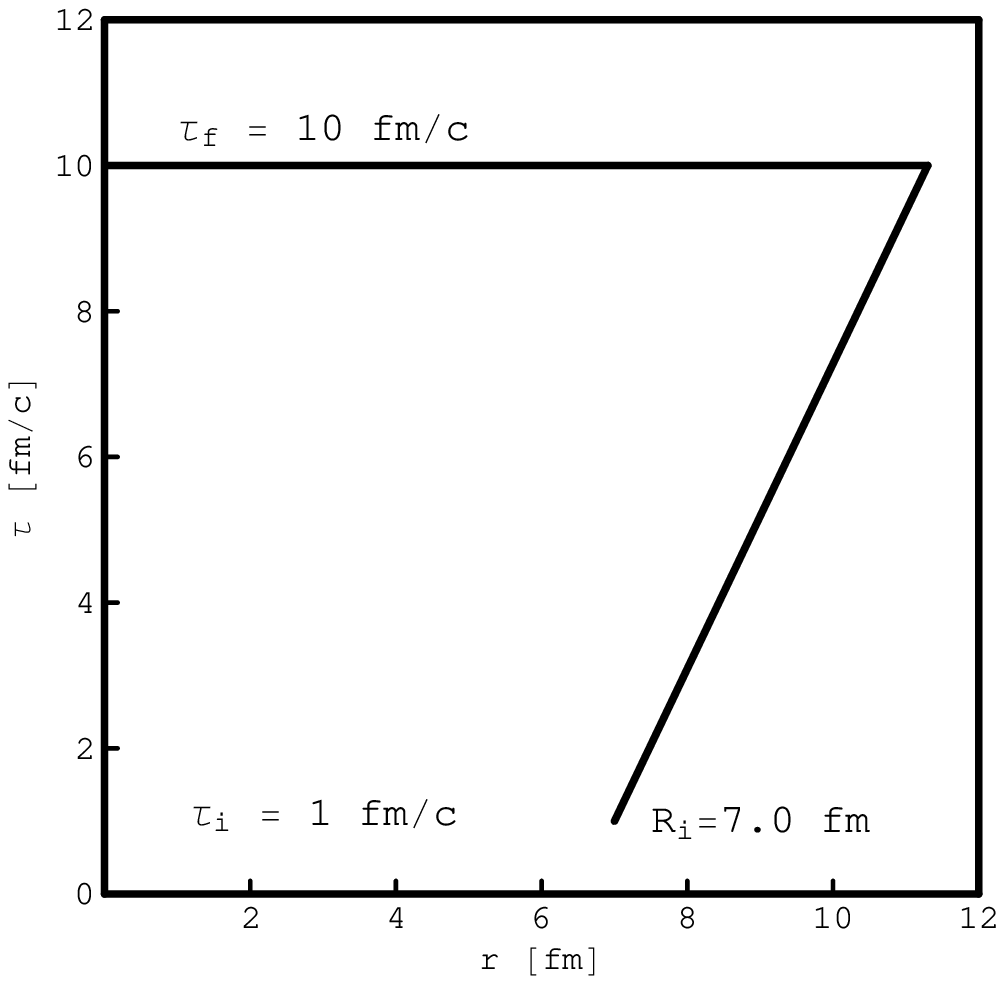}
\caption{ The enclosed freeze-out hypersurface $\tau(r)$. See the
text for details. } \label{fig1}
\end{figure}

\newpage

\begin{figure}[h]
\centering
\includegraphics[scale=0.8]{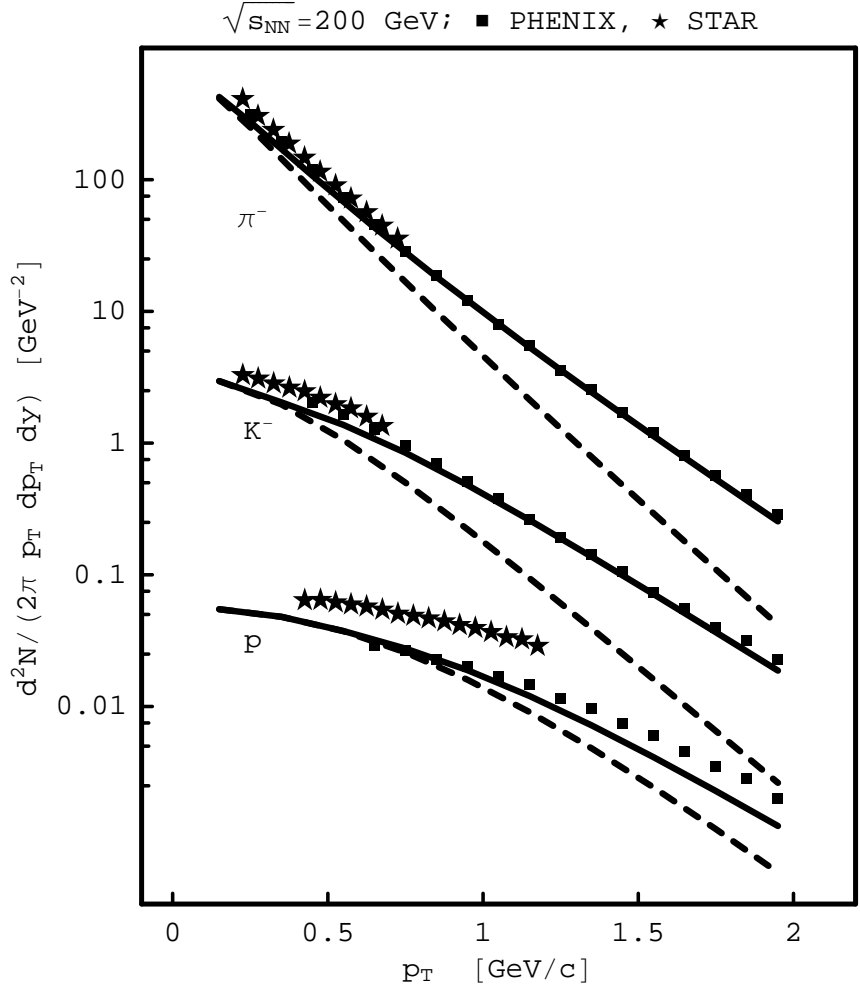}
\caption{ Comparison of the single-particle momentum spectra
measured by the PHENIX Collaboration
  with the model calculations performed
 for whole enclosed hypersurface (solid lines) and for  shelf-like part, see Fig. 1, solely (dashed lines).
 For convenience of plots location the measured $p$ spectra are reduced in $100$
 times and $K^{-}$ spectra in 8 times. The STAR data are  not fitted  and are presented for illustration.} \label{fig2}
\end{figure}

\newpage

\begin{figure}[h]
\centering
\includegraphics[scale=0.9]{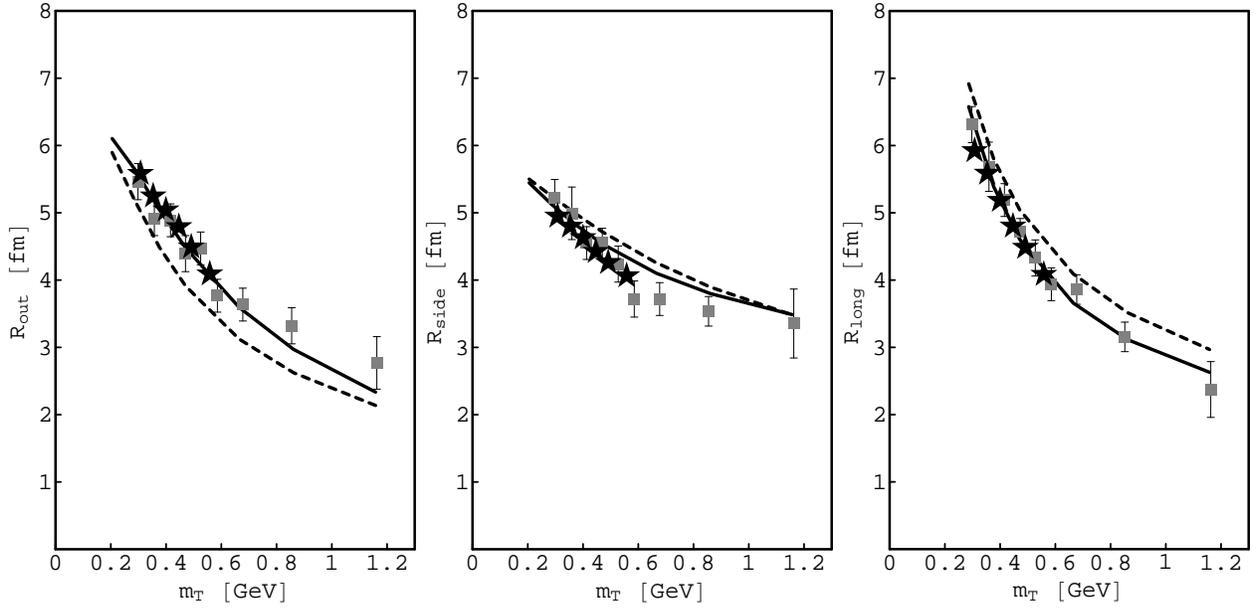}
\caption{ Comparison of the pion source $R_{out}$, $R_{side}$,
$R_{long}$ radii measured by the STAR (stars) and PHENIX (boxes)
Collaborations  with model calculations performed for whole enclosed
hypersurface (solid lines) and for  shelf-like part
 only (dashed lines). The PHENIX data are adjusted
for most central bin (see the text for details). } \label{fig3}
\end{figure}

\newpage

\begin{figure}[h]
\centering
\includegraphics[scale=1.4]{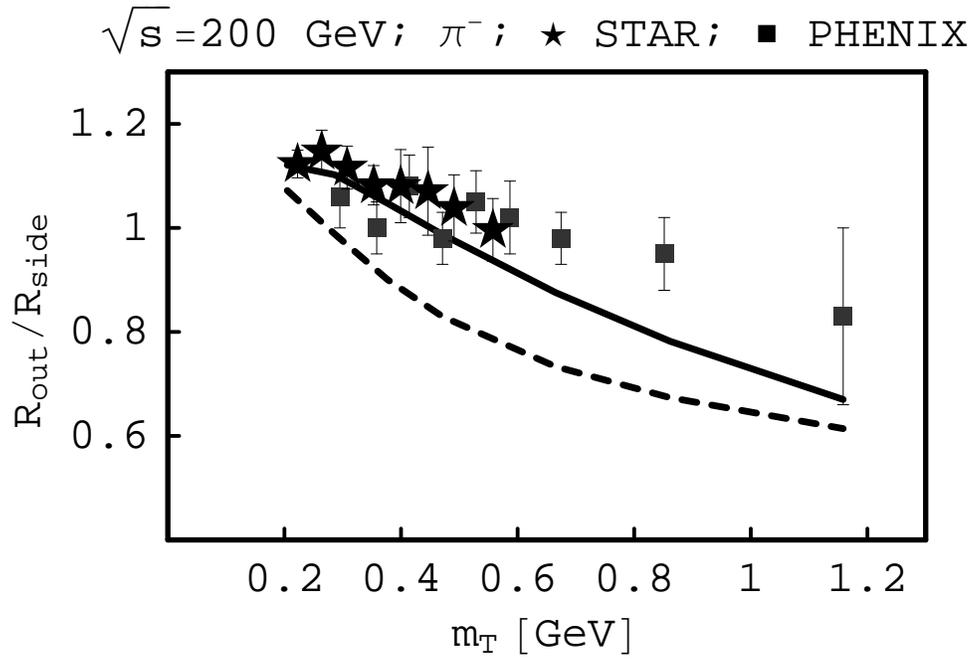}
\caption{ Comparison of the measured by the STAR (stars) and PHENIX (boxes) Collaborations pion $R_{out}/R_{side}$ ratios
with  model calculations performed for whole enclosed hypersurface (solid line) and for  shelf-like part only
   (dashed line).} \label{fig6}
\end{figure}

\end{document}